%%%%%%%%%%%%%%%%%%%%%%%%%%%%%%%%%%%%%%%%%%%%%%%%%
%    Finite Decoherence Factor
%%%%%%%%%%%%%%%%%%%%%%%%%%%%%%%%%%%%%%%%%%%%%%%%%
\input phyzzx
%%%%%%%%%%%%%%%%%  Macros  %%%%%%%%%%%%%%%%%%%%%%
\def\half{{1 \over 2}}

\def\parderi#1#2{{\partial #1\over\partial #2}}
\def\deri#1{\partial_#1}
\mathchardef\Lag="724C
\def\sqr#1#2{{\vcenter{\hrule height.#2pt
      \hbox{\vrule width.#2pt height#1pt \kern#1pt
          \vrule width.#2pt}
      \hrule height.#2pt}}}
\def\square{{\mathchoice{\sqr84}{\sqr84}{\sqr{5.0}3}{\sqr{3.5}3}}}

\def\omegan{\omega_n}
\def\fn{f_n}

\def\Prod#1{\mathop{\Pi}_#1}
\def\bfG{{\bf G}}
%%%%%%%%%%%%%%  Title page  %%%%%%%%%%%%%%%%%%%%%
\pubnum={TIT/HEP-313/COSMO-63}

\titlepage

\title{On Divergence of Decoherence Factor \nextline
           in Quantum Cosmology
}
\author{Takashi Okamura}
\address{Department of Physics, Tokyo Institute of Technology, \nextline
Oh-okayama, Meguroku, Tokyo 152, Japan 
}

\abstract{
To discuss the quantum to classical transition in quantum cosmology,
we study the decoherence factor and the peak of the Wigner function,
which respectively represent the degree of decoherence and 
the degree to which the classical motion of the Universe is defined, 
in a Robertson-Walker universe model coupled with a scalar field.
It is known that the decoherence factor is divergent in some cases.
This implies that perfect decoherence occurs, and 
classical correlation criterion fails.
In this paper we discuss the divergence of the decoherence factor
in some detail and obtain the constraints for decoherence factor
to be convergent, making use of the arbitrariness defining 
the reduced density matrix.
The result 
%that the choice of conformal rescaled field 
%as environment and conformal coupling make the decoherence factor finite
is discussed in connection with the arbitrariness of 
{\it system/environment} splitting.
}

\endpage
%%%%%%%%%%%%%%%%%%%%%%%%%%%%%%%%%%%%%%%%%%%%%%%%%
\chapter{Introduction}
%%%%%%%%%%%%%%%%%%%%%%%%%%%%%%%%%%%%%%%%%%%%%%%%%
Many researchers accept that quantum theory can be applied to
all events and even to the entire Universe.
However, macroscopic objects, including our universe, behave
classically. If we seriously consider quantum theory as a universal 
theory, it is a crucial problem to derive classical behavior from
quantum theory under some suitable conditions.

To characterize classicality, two criteria are generally used 
as fundamental, classical correlation and quantum decoherence. 

Strong correlation is necessary to predict from the wave function of 
the Universe.
It is a reason that we cannot have probabilistic predictions 
from the wave function of the Universe that is a single system 
and can only predict the event whose (conditional) probability is 
almost unity, $\ie$, the strong peak of the wave function 
or Wigner function.
\REFS\geroch{
	R.Geroch, Nous {\bf 18}(1984),617.
}
\REFSCON\hartle{
	J.B.Hartle, in {\it Gravitation in Astrophysics}, 
	proceedings of the NATO Advanced Study Institute, Cargese, France,
	1986, edited by B.Carter and J.Hartle 
	(NATO ASI Series B:Physics, Vol.156)(Plenum, New York, 1987).
}\refsend
The existence of correlation between configurations and momenta 
along a classical trajectory
implies that the system follows classical equations of motion.
In order to analyze whether a given wave function has 
classical correlation, we often examine the peak of the Wigner function.
\Ref\halliwellwigner{
	J.J.Halliwell, Phys.Rev.{\bf D36}(1987),3626.
}

We may expect that the WKB state is classical in a sense and 
that the Wigner function associated with it shows classical correlation.
However, the Wigner function of the WKB state 
not only has no peak around the classical trajectory in phase space 
but also oscillates very rapidly. 
\REFS\habib{
	S.Habib, Phys.Rev.{\bf D42}(1990),2566.
}
\REFSCON\berry{
	M.V.Berry, Philos. Trans. R. Soc.{\bf 287}(1977),237.
}\refsend
The cause of these phenomena is the existence of quantum interference.
To acquire classical correlation, we need some coarse graining.
\Ref\habiblaflamme{
	S.Habib and R.Laflamme, Phy.Rev.{\bf D42}(1990),4056.
}

Quantum decoherence induces the coarse graining necessary for 
the Wigner function to have a peak.
\refmark\habiblaflamme
And originally, quantum decoherence is necessary for individual systems 
in a composite system to have definite characters, 
\Ref\despagnat{
	B.d'Espagnat, {\it Conceptual Foundations of Quantum Mechanics}
	(Benjamin, Menlo Park, Calif., 1971).
}
and necessary to assign the probability into each history of the system.
\Ref\gellmannhartle{
	M.Gell-Mann and J.B.Hartle, in {\it Comlexity, Entropy and the
	Physics of Information}, edited by W.H.Zurek,
	Santa Fe Institute Studies in the Sciences of Complexity Vol.VIII
	(Addison-Wesley, Reading, MA, 1990).
}

Classical correlation and quantum decoherence are not only complementary
but also exclusive in part,
because decoherence induces a coarse graining on 
the Wigner function, and the coarse graining leads to 
a spreading of the Wigner function.
\refmark\habiblaflamme
\Ref\morikawa{
	M.Morikawa, Phys.Rev.{\bf D42}(1990),2929.
}
Let us assume that we have a superposed state 
in the position representation, 
and after the decoherence mechanism acts, its density matrix 
associated with the state is diagonalized in that representation.
Therefore we have a classical ensemble consisting of the states whose
positions are decided.
Each state is a position eigenstate so that the conjugate momentum is
indefinite.
So we have no classical correlation 
between positions and momenta in phase space.
To obtain a definite classical correlation, 
decoherence should be moderate.

A mechanism leading to decoherence is 
the ``environment induced superselection rule''.
\Ref\zurek{
	W.H.Zurek, Phys.Rev.{\bf D24}(1981),1516;{\bf 26}(1982),1862.
}
Macroscopic objects are not isolated from their environment.
We finally need reduced information because we cannot observe 
the entire Universe.
This inevitable interaction with the environment induces non-unitary
evolution on the relevant degrees of freedom called the {\it system}.

This mechanism has been applied to quantum cosmology.
\REFS\halliwellII{
	J.J.Halliwell, Phys.Rev.{\bf D39}(1989),2912.
}
\REFSCON\kiefer{
	C.Kiefer, Class.Quantum.Grav.{\bf 4}(1987),1369.
}
\REFSCON\padmanabhan{
	T.Padmanabhan, Phys.Rev.{\bf D39}(1989),2924.
}
\REFSCON\morikawaII{
	M.Morikawa, Phys.Rev.{\bf D40}(1989),4023.
}
\REFSCON\pazsinhaI{
	J.P.Paz and S.Sinha, Phys.Rev.{\bf D44}(1991),1038.
}
\REFSCON\pazsinhaII{
	J.P.Paz and S.Sinha, Phys.Rev.{\bf D45}(1992),2823.
}\refsend
There, decoherence of the spacetime and classical correlation, $\ie$,
derivation of the semiclassical Einstein equation, has been discussed.
Decoherence and correlation of the scale factor of the 
Robertson-Walker universe 
by an inhomogeneous scalar field as environment
are usually examined under the WKB ansatz, 
and the degree of decoherence is expressed by the overlap integral, 
called the decoherence factor, 
between the states of scalar fields which have evolved 
on different WKB histories.

In some cosmological models, we face perfect decoherence of
the scale factor.
\refmark\kiefer\refmark\pazsinhaI\refmark\pazsinhaII
Because we deal with a system of an infinite numbers of 
degrees of freedom, we essentially
need regularization and renormalization procedures.
However, this feature is not removed by the standard renormalization 
procedure.
\refmark\pazsinhaII
Further this is an undesirable feature for classical correlation.
So we usually introduce a cutoff to 
the degrees of freedom of scalar modes as the environment.
Unfortunately, whether the scale factor behaves classically is strongly
dependent  on the choice of cutoff.
\refmark\halliwellII\refmark\kiefer\refmark\habiblaflamme
Meanwhile, we have a model that shows moderate decoherence.
Several authors related the cause of perfect decoherence to 
the lack of adiabaticity of the WKB histories.
\refmark\pazsinhaI\refmark\pazsinhaII

We always have arbitrariness in defining the reduced density matrix.
\Ref\laflammelouko{
	R.Laflamme and J.Louko, Phys.Rev.{\bf D43}(1991),3317.
}
The reduced density matrix may be defined by integrating out the modes
$\{\phi_n\}$ as the environment,
$$
	\rho_{red}(a,a')=\int [\Prod n d\phi_n]~
	\rho(a,\{\phi_n\};a',\{\phi_n\})~.
$$
Alternatively, we may define it by integrating out $\{f_n=ag(a)\phi_n\}$
as the environment,
$$
	\bar\rho_{red}(a,a')=\int 
	[\Prod n df_n (a g(a))^{-1/2}(a' g(a'))^{-1/2}]~
	\rho(a,\{{f_n \over a g(a)}\};a',\{{f_n \over a' g(a')}\})~.
$$
The arbitrariness of the choice of the environment variables 
brings change to the conjugate momenta 
so that the portion spanned by the Hilbert space of the system 
within the total Hilbert space changes.
Except for the coincidence limit, $a=a'$, 
$\rho_{red} \neq \bar\rho_{red}$.
Then, is it a natural choice which reduced density matrix we use for 
the decoherence argument of the scale factor?
Laflamme and Louko 
\refmark\laflammelouko
assert that the choice $g(a)=1$ is preferable to
$g(a)=a^{-1}$ because the decoherence induced by the former is more
directly related to particle production.

In this paper, through a model consisting of a scale factor 
and free scalar field with curvature coupling, 
we somewhat carefully discuss the question of which 
which choice for reduced density matrix leads to moderate decoherence.
The paper is organized as follows.
In Sec.II we define the decoherence factor induced by integrating out
the rescaled scalar field according to the usual procedure.
In Sec.III we relate the variance of Gaussian solution 
for the Schr\"odinger equation of rescaled scalar field 
to mode functions that are solutions of Klein-Gordon equation, 
and review the known results briefly.
In Sec.4, using the constraints on large frequency behavior of 
the mode function by Bogoliubov implementability,
we obtain the constraints on rescaling $g(a)$ for the scalar field 
and the curvature coupling constant.
Sec.V is devoted to a summary.
%%%%%%%%%%%%%%%%%%%%%%%%%%%%%%%%%%%%%%%%%%%%%%%%%
\chapter{Decoherence Functional}
%%%%%%%%%%%%%%%%%%%%%%%%%%%%%%%%%%%%%%%%%%%%%%%%%
We consider the Robertson-Walker universe with a free scalar field, 
and examine the decoherence of the scale factor 
by using the rescaled field as the environment.
The Lagrangian density of this system is 
$$
	\Lag={M^2 \over 12}\sqrt{-g}~R-{\sqrt{-g} \over 2}
	[(\nabla \Phi)^2+m^2\Phi^2+\xi R\Phi^2]~.
\eqn\Lagrangiandensity
$$
We choose rescaled modes $\{f_n\}$ as the environment,
$$
	\Phi(t,\vec x)=\sum_n {\fn(t) \over a g(a)} Y_n(\vec x)~,
\eqn\phidef
$$
where $\{ Y_n \}$ are normalized eigenfunctions of the Laplacian 
with respect to $h_{ij}$, and the index $n$ is an abstract index 
classifying the eigenfunctions $Y_n$.
The eigenvalue of $Y_n$ is $-n^2=-(k^2-K)$.

Using the RW metric,
$$\eqalign{
	ds^2&=-N(t)^2 dt^2+a^2(t) h_{ij}(\vec x)dx^i dx^j \cr
	&=-N^2 dt^2+a^2 \Bigl({dr^2 \over 1-Kr^2}+r^2 d\Omega^2 \Bigr)~,
\cr}\eqn\metric
$$
the Lagrangian is written as
$$\eqalignno{
	L&={1 \over 2N}G_{\mu\nu}\dot X^\mu \dot X^\nu-NU~,\qquad\qquad
	X^\mu=(a,\{ \fn \} )~,
&\eqnalign\LagranI\cr
	U(X)&=-M^2{Ka \over 2}+{n^2+m^2a^2+6\xi K \over 2a g^2}\fn^2 \cr
	    &\equiv M^2 V(a)+{\tilde\omegan^2 \over 2 a g^2}f_n^2~,
&\eqnalign\potentialI\cr
	G_{\mu\nu}&=\pmatrix{
	        \bfG_{aa}\Bigl[F^{-1}-\sum_n 
	\Bigl({M a \lambda_n \over g}\Bigr)^2 \Bigr]
              & {M^2 a^2 \over g^2}\lambda_n \cr
              {M^2 a^2 \over g^2}\lambda_{\bar n} & 
	{a \over g^2}\delta_{\bar n n} \cr  }~,
&\eqnalign\supermlower\cr
	\bfG_{aa}& = -M^2 a~, \qquad\qquad \bfG^{aa}=-{1 \over M^2 a}~,
&\eqnalign\bgsuperm\cr
	\lambda_n&=\bfG^{aa}\Bigl(\ln {g \over a^{6\xi-1}}\Bigr)' \fn~,
&\eqnalign\deflambdan\cr
	F&=\Bigl[1-{6\xi(1-6\xi) \over M^2a^2g^2(a)}
	\sum_n \fn^2 \Bigr]^{-1}~,
&\eqnalign\defF\cr
}$$
where $\bfG_{aa}$ is the supermetric in the case of no coupling to 
the scalar field, and the dash represents the derivative 
with respect to the scale factor $a$.

The Hamiltonian of this system becomes
$$\eqalignno{
	H&=\half G^{\mu\nu}P_\mu P_\nu+U~,
&\eqnalign\hamiltonian\cr
	G^{\mu\nu}&=\pmatrix{
	\bfG^{aa}F  &  F \lambda_n \cr
	F \lambda_{\bar n} & 
	{g^2 \over a}\delta_{\bar n n}+F\bfG_{aa}\lambda_{\bar n}\lambda_n 
	\cr}~.
&\eqnalign\smetric\cr
}$$

Quantizing this system, we obtain the Wheeler-DeWitt equation,
$$
	[-\half \square_G + U]\Psi(X^\mu)=0~,
\eqn\WDeq
$$
where $\square_G$ is the D'Alembertian 
with respect to supermetric $G_{\mu\nu}$, and the integral measure for
the square of wave functions is $da \Prod n d\fn \sqrt{-G}\sim 
da \sqrt{-\bfG_{aa}}\Prod n [d\fn (a^{1/2}/g)]$.
Due to the existence of the Planck scale, we look for the solutions 
under the Born-Oppenheimer approximation,
$$\eqalign{
	\Psi(a,\{\fn\})&=C(a)e^{iS(a)}\chi(\{\fn\}~;a)~,
\cr
	O(S(a)) &\sim O(M^2)~.
}\eqn\WKB
$$
Substituting Eq.\WKB~ into Eq.\WDeq~ and expanding in powers of $M^{-2}$,
we obtain the Hamilton-Jacobi equation from leading term,
$$
	\half \bfG^{aa} (\deri a S)^2+ M^2 V=0~,
\eqn\HJeq
$$
To next order in $M^{-2}$, 
we obtain the probability conservation equation
and the Schr\"odinger equation,
$$\eqalignno{
	&\deri a(\sqrt{-\bfG}\bfG^{aa} C^2 \deri a S)=0~,
&\eqnalign\probcons\cr
 &i(\bfG^{aa}\deri aS)(\deri a+\half \deri a\ln{a^{1/2} \over g})\chi
\cr&\qquad
     =\Bigl[-{g^2 \over 2 a}\partial^2_n
	-i{\deri aS \over 2}(\lambda_n\deri n+\deri n\lambda_n)
	+{\omega^2_n \over 2a g^2}\fn^2 \Bigr]\chi~,
&\eqnalign\SchI\cr
	&\omega^2_n=\tilde\omega_n^2-6\xi(1-6\xi)(\bfG^{aa}\deri aS)^2~,
&\eqnalign\litleomega\cr
}$$
where we choose the prefactor $C$ to be WKB prefactor in the case of 
the pure RW universe. By this choice of $C$, 
wave function of the scalar field is normalized as
$$
	1=\int \Prod n \Bigl\{ d\fn {a^{1/2} \over g} \Bigr\}
	|\chi|^2~.
\eqn\normalized
$$

We define WKB time as
$$
	{1 \over N}\parderi{}{t}\equiv (\bfG^{aa}\deri aS)\deri a~,
$$
so $\bfG^{aa}\deri aS=\dot a/N$.
Eq.\SchI~ is rewritten as
$$\eqalignno{
	{1 \over N}\parderi \psi t &=\Bigl[ -{g^2 \over 2 a}{\deri n}^2
	+{1 \over 2a g^2}\Omega_n^2\fn^2 \Bigr]\psi~,
&\eqnalign\SchII\cr
	\psi&=\Bigl({a^{1/2} \over g}\Bigr)^{1/2}
  \exp\Bigl\{ {i a \over 2 Ng^2}\Bigl(\ln{g \over a^{6\xi-1}}\Bigr)^\cdot
	\fn^2 \Bigr\}\chi~,
&\eqnalign\defpsi\cr
	\Omega_n^2&=k^2-K+a^2(m^2+(\xi-{1 \over 6}) R)-{a \over N}
	\Bigl({a\dot g \over Ng} \Bigr)^\cdot
	+\Bigl({a\dot g \over Ng} \Bigr)^2 
\cr
	& \equiv \nu_n^2-{a \over N}\Bigl({a\dot g \over Ng} \Bigr)^\cdot
	+\Bigl({a\dot g \over Ng} \Bigr)^2~.
&\eqnalign\defOmega\cr
}$$

In quantum mechanics, superposed states can also exist, and the general
form of wave function is
$$
	\Psi=\sum_l C_l(a)\exp[iS_l(a)]\chi_l(\{\fn\}~;a)~,
\eqn\superposed
$$
where the index $l$ is a parameter 
that classifies the solutions of Eq.\HJeq~.
In our case, $l=\pm$ where $+$($-$) corresponds to 
expanding(contracting) universe, $S_-=-S_+$, and 
$\partial / N \partial t_-=- \partial / N \partial t_+$.
Hereafter we use a "dot" to represent the derivative 
with respect to $t_+$.

We define the density matrix of the total system and
the reduced density matrix of the scale factor.
First, we define the total density matrix as
$$
	\rho(a,\{ \fn \}~;a',\{ f'_n \}~)
	=\Psi(a,\{\fn\}~)\Psi^*(a',\{f'_n \}~)~.
\eqn\totaldensitymatrix
$$
From the bi-scalar character of the density matrix, we define 
the reduced density matrix as
$$
	\rho_{red}(a,a')=\int \Prod n \Bigl\{ 
	d\fn \Bigl({a^{1/2} \over g(a)}\Bigr)^{1/2}
	\Bigl({a'^{1/2} \over g(a')}\Bigr)^{1/2} \Bigr\}
	\rho(a,\{\fn\}~;a',\{\fn \}~)~.
\eqn\reduced
$$
From Eq.\superposed~, we obtain
$$\eqalignno{
	&\rho_{red}(a,a')=\sum_{l,l'}\rho^{(0)}_{l,l'}(a,a')I_{l,l'}(a,a')~,
&\eqnalign\reducedII\cr
 &\rho^{(0)}_{l,l'}(a,a')=C_l(a)C^*_{l'}(a')\exp[i(S_l(a)-S_{l'}(a'))]~,
&\eqnalign\purerho\cr
	&I_{l,l'}(a,a')=\int \Prod n d\fn 
	\exp[-{i \over 2}(\theta_l-\theta'_{l'})\fn^2]
	\psi_l(\{\fn\}~;a) \psi^*_{l'}(\{\fn\}~;a')~,
&\eqnalign\DF\cr
	&\theta_l(a)
	={l a \over N g^2}\Bigl(\ln{g \over a^{6\xi-1}}\Bigr)^\cdot~,
&\eqnalign\thetal\cr
}$$
where $\rho^{(0)}$ is the density matrix 
in the case of a pure RW universe, 
and $I$ is often called the decoherence factor.
From normalization, Eq.\normalized~, at the coincidence limit, $l=l'$ 
and $a=a'$, we obtain $|I_{l,l}(a,a)|=1$.
Further, by the Cauchy-Schwartz inequality 
we generally obtain the inequality $|I_{l,l'}(a,a')| \le 1$.
Thus, the decoherence factor reflects 
the degree of decoherence of the scale factor.
%%%%%%%%%%%%%%%%%%%%%%%%%%%%%%%%%%%%%%%%%%%%%%%%%
\chapter{Solution of the Schr\"odinger equation and Known Results}
%%%%%%%%%%%%%%%%%%%%%%%%%%%%%%%%%%%%%%%%%%%%%%%%%
As we discussed in the previous section, to examine decoherence 
under the WKB ansatz, we first solve the Hamilton-Jacobi equation, 
Eq.\HJeq~, for the background geometry, and 
next solve the Schr\"odinger equation, Eq.\SchII~, 
in the background geometry 
that is a solution of the Hamilton-Jacobi equation.
Finally, we estimate the decoherence factor, Eq.\DF~.

Because there is no mode coupling in Eq.\SchII~, we deal with each mode 
separately.
% and often omit index $l$.
Eq.\SchII~ has a quadratic form, 
and we seek solutions of the Gaussian form
$$
	\psi_{nl}=\Bigl({\bar B_{nlR} \over \pi}\Bigr)^{1/4}
	\exp[iA_{nl}(t)-\half \bar B_{nl}(t) \fn^2]~,
\eqn\gaussian
$$
where $\bar B_{nlR}=\hbox{Re}(\bar B_{nl})>0$.
From the Schr\"odinger equation Eq.\SchII~, we obtain
$$\eqalignno{
	&l{\dot A_{nl} \over N}=-{g^2 \over 2 a}\bar B_{nlR}~,
&\eqnalign\dotA\cr
	&l{\dot{\bar B}_{nl} \over N}=-i{g^2 \over a}\bar B_{nl}^2
	+{i \over a g^2}\Omega_n^2~.
&\eqnalign\dotB\cr
}$$
We introduce the variable $u_{nl}$ as
$$
	\bar B_{nl}=-i{l a \over N g^2}[\ln u_{nl}^* + \ln g]^\cdot~,
\eqn\introu
$$
and Eq.\dotB~ is rewritten as
$$\eqalign{
	&{a \over N}\Bigl( {a \over N}\dot u_{nl} \Bigr)^\cdot 
	+\nu^2_n u_{nl}=0~,
\cr
	&\nu_n^2=k^2+a^2[m^2+(\xi-{1 \over 6})R]~.
\cr
}\eqn\modeeq
$$
Eq.\modeeq~ is just the Klein-Gordon equation.
So $u_{nl}$ is a mode function, and the choice of the mode function 
corresponds to the initial condition of the wave function, $\psi_{nl}$.

In the Gaussian form, Eq.\gaussian~, the decoherence factor, Eq.\DF~ is
written as
$$\eqalignno{
	&I_{l,l'}(a,a')=\Prod n \exp[i(A_{nl}(a)-A_{nl'}(a'))]\Bigl[
	{4 B_{nlR}(a) B_{nl'R}(a') \over (B_{nl}(a)+B^*_{nl'}(a'))^2}
	\Bigr]^{1/4}~,
&\eqnalign\DFII\cr
	&B_{nl}(a)=\bar B_{nl}(a)+i~\theta_l(a)
	=-i{l a \over N g^2}[\ln u_{nl}^*+(6\xi-1)\ln a]^\cdot~.
&\eqnalign\defB\cr
}$$

Let us try to apply the procedures described above to some models. 
First, we consider the case of a massless minimally coupled field 
on the deSitter background as a zeroth order WKB solution.
\refmark\halliwellII
If we choose the deSitter invariant vacuum 
as the boundary condition of the mode functions,
the functions $B_{nl}$ are given by
$$
	B_{nl}(a)=(k^2-1)a^2
	{k+i~ l~(H^2 a^2-1)^{1/2} \over k^2+ H^2 a^2-1}~,
\eqn\eq
$$
where we chose the rescaling $g(a)=a^{-1}$.
So we obtain
\refmark\halliwellII
$$
	|I_{l,l'}(a,a')|=\mathop{\Pi}_{k=2} \Bigl[ 
	1+{(a^2-a'^2)^2 \over 4 a^2 a'^2}+
	{(a'^2 l\sqrt{H^2 a^2-1}-a^2 l' \sqrt{H^2 a'^2-1}~)^2 
	\over 4k^2 a^2 a'^2}
	\Bigr]^{-k^2/4}~.
\eqn\DFdesitter
$$
By calculating the infinite product, we obtain perfect decoherence 
between the expanding universe and contracting one, 
and perfect decoherence within a WKB branch,
$$\eqalign{
	&|I_{+-}(a,a)|=\mathop{\Pi}_{k=2}^\infty
	\Bigl[ 1+{H^2 a^2-1 \over k^2} \Bigr]^{-k^2/4}=0~,
\cr
	&|I_{l,l}(a,a')|< \mathop{\Pi}_{k=2}^\infty
	\Bigl[ 1+{(a^2-a'^2)^2 \over 4 a^2 a'^2} \Bigr]^{-k^2/4}=0~
	\qquad \hbox{for } a \ne a'~.
\cr}\eqn\eq
$$

Next we consider the case of a massive conformally coupled field
on an adiabatically evolving universe.
\refmark\pazsinhaI\refmark\pazsinhaII
\Ref\calzettamazzitelli{
	E.Calzetta and F.Mazzitelli, Phys.Rev.{\bf D42}(1990),4066.
}
Let us set, at scale factor $a=a_0$, 
the initial condition of the scalar field as in-vacuum
and choose the rescaling $g(a)=1$ and $N=a$.
Using Bogoliubov coefficients, which are estimated by 
a standard WKB technique, 
the in-vacuum is related to the adiabatic out-vacuum as
$$\eqalign{
	&u_n^{in}=\alpha_n u_n^{out}+\beta_n {u_n^{out}}^*~,
\cr
	&u_n^{out}\sim {\exp(-i \int^\eta \nu_n d\eta') \over
	              (2 \nu_n)^{1/2} }~,
\cr
	&\beta_n={i \over 2}\exp[-{\gamma \over 2}(k^2+m^2 a_0^2)]~,
	\qquad \gamma={\pi \over \sqrt{m^2 a_0 V'(a_0)}}~,
\cr
	&\alpha_n \sim 1~.
\cr}\eqn\inout
$$
Thus
$$
	B_{n+}(a)=\nu_n \Bigl[ {\alpha^*_n - \beta^*_n u_n/u_n^{out*} 
	\over \alpha^*_n + \beta^*_n u_n^{out}/u_n^{out*} } \Bigr]
	+ i{\dot \nu_n \over 2 \nu_n}=B^*_{n-}(a)~.
\eqn\eq
$$
The norm of the decoherence factor between the expanding universe and 
contracting universe is given by
\refmark\calzettamazzitelli\refmark\pazsinhaII
$$
	|I_{+-}(a,a)|=\mathop{\Pi}_{k=2}^\infty
	\Bigl[ 1-{1 \over \nu_n^2}
	\Bigl\{ 2|\beta_n|\cos(2\int^\eta \nu_n)-{\dot\nu_n \over 2\nu_n^2}
	\Bigr\}^2 \Bigr]^{k^2/4}~,
\eqn\exadiabatic
$$
and due to $O(\nu_n^{-2})=O(\dot\nu_n/\nu_n^2)=O(k^{-2})$, 
the infinite product on the R.H.S. of Eq.\exadiabatic~ is convergent,
so that moderate decoherence results.
Similarly the decoherence factor within a WKB branch is also calculated,
and it does not vanish.
\refmark\pazsinhaI
%
%%%%%%%%%%%%%%%%%%%%%%%%%%%%%%%%%%%%%%%%%%%%%%%%%
\chapter{Constraints on rescaling and curvature coupling constant}
%%%%%%%%%%%%%%%%%%%%%%%%%%%%%%%%%%%%%%%%%%%%%%%%%
We would like to know how the scale factor decoheres.
Thus we concentrate on the norm of the decoherence factor,
$$\eqalign{
	|I_{l,l'}(a,a')|&=\Prod n \Bigl[ 
	{4B_{nlR}(a)B_{nl'R}(a') \over |B_{nl}(a)+B^*_{nl'}(a')|^2}
	\Bigr]^{1/4} \cr
	&=\Prod n \Bigl[1-
	{|B_{nl}(a)-B_{nl'}(a')|^2 \over |B_{nl}(a)+B^*_{nl'}(a')|^2}
	\Bigr]^{1/4}~. \cr
}\eqn\normDF
$$
By virtue of a well known theorem for infinite products,
the divergence of Eq.\normDF~ is equivalent to that of 
$$
	\sum_n {|B_{nl}(a)-B_{nl'}(a')|^2 \over |B_{nl}(a)+B^*_{nl'}(a')|^2}~.
\eqn\DFsum
$$
In our case, in the large $k$ limit we can replace 
the summantion for mode $k$ into an integral over $k$:
$$
	\sum_n \quad \longrightarrow \quad \int dk k^2~.
$$
If Eq.\DFsum~ is convergent, it must hold that for large $k$,
$$
	{|B_{nl}(a)-B_{nl'}(a')|^2 \over |B_{nl}(a)+B^*_{nl'}(a')|^2}
	\sim o(k^{-3})~.
\eqn\orderDFsum
$$

In order to obtain the constraints for the decoherence factor 
to be finite, we attach some physically reasonable conditions 
to the mode functions.
We expect that at each time we can have a particle picture and that
the Fock representation set at each time can be related 
unitary equivalently, $\ie$, Bogoliubov implementability.
\REFS\fulling{
	S.A.Fulling, Gen.Rel.Grav.{\bf 10}(1979),807.
}
\REFSCON\kodama{
	H.Kodama, Prog.Theor.Phys.{\bf 65}(1981),507.
}

This Bogoliubov implementability gives some constraints on 
the behavior of the mode function in the large $k$ limit:
\refmark\kodama
$$\eqalignno{
	\mu_{nl} &\equiv \hbox{Re}
	\Bigl[-i{l a \over N}{ \dot u_{nl}^* \over u_{nl}^* }\Bigr]
&\eqnalign\defmu\cr
	& \sim \nu_n+o(k^{-1/2})~,
&\eqnalign\behaviormu\cr
	\gamma_n &\equiv -\hbox{Im}
	\Bigl[-i{l a \over N}{\dot u_{nl}^* \over u_{nl}^*}\Bigr]
&\eqnalign\defgamma\cr
	&\sim o(k^{-1/2})~.
&\eqnalign\behaviorgamma\cr
}$$
For example, the behavior of each Eqs.\defmu~ and \defgamma~ 
in the large $k$ limit for adiabatic vacuume is, respectively,
$$\eqalign{
	&\mu_{nl} =\nu_n+o(k^{-2})~,
\cr
	&\gamma_{nl}=o(k^{-2})~,
\cr}\eqn\adiabatic
$$
so that Eqs.\behaviormu~ and \behaviorgamma~ hold.
For the Hamiltonian diagonalization vacuum,
the behavior of Eq.\defgamma~ in the large $k$ limit is
$$
	\gamma_{nl}=(1-6 \xi){l \dot a \over Na}~,
\eqn\Hdv
$$
so that Eq.\behaviorgamma~ does not hold, except for when $\xi=1/6$.
\refmark\fulling\refmark\kodama

Under the WKB ansatz, Eq.\superposed~, we take the constraints 
Eqs.\behaviormu~ and \behaviorgamma~ for granted 
because the back reaction correction to the background evolution 
should be a small finite quantity
so that the particle production is finite.

Due to the constraints \behaviormu~ and \behaviorgamma~, 
$$
	B_{nl}={1 \over g^2}\Bigl[\mu_{nl}-i\gamma_{nl}
	+i{l a \over N}(\ln a^{1-6\xi})^\cdot \Bigr]\sim O(k)~,
\eqn\orderB
$$
and thus, from Eq.\orderDFsum~, we obtain the constraint
$$
	|B_{nl}(a)-B_{nl'}(a')|^2 \sim o(k^{-1})~.
\eqn\orderdiffB
$$
From Eq.\orderdiffB~, we derive the constraints on rescaling $g(a)$ and 
the curvature coupling constant to obtain moderate decoherence.

%--------------------------------------------
\section{decoherence between different WKB branches}
%--------------------------------------------
As different WKB branches in our model, we have the expanding universe
and contracting universe.
First, we examine the decoherence between these two states.
Its degree is measured by
$$
	|I_{+-}(a,a)|=\Prod n \Bigl[1-
	{|B_{n+}(a)-B_{n-}(a)|^2 \over |B_{n+}(a)+B^*_{n-}(a)|^2}
	\Bigr]^{1/4}~,
\eqn\diffWKB
$$
and its convergence is equivalent to Eq.\orderdiffB~ with $a'=a$:
$$
	|B_{n+}(a)-B_{n-}(a)|^2={1 \over g^4}\Bigl[
	(\mu_{n+}-\mu_{n-})^2+\{ \gamma_{n+}-\gamma_{n-}-
	2i{a \over N}(\ln a^{1-6\xi})^\cdot \}^2 \Bigr] \sim o(k^{-1})~.
\eqn\diffWKBII
$$
Under Eqs.\behaviormu and \behaviorgamma~, for the decoherence factor, 
Eq.\diffWKB~, to be finite, we obtain the constraint 
on the curvature coupling constant
$$
	\xi={1 \over 6}~,
\eqn\eq
$$
$\ie$, conformal coupling.
%--------------------------------------------
\section{decoherence within a WKB branch}
%--------------------------------------------
Next we examine decoherence of the scale factor within a WKB branch.
We omit the index $l$ in the decoherence factor, and so on.
The degree of decoherence is measured by
$$
	|I(a,a')|=\Prod n \Bigl[1-
	{|B_n(a)-B_n(a')|^2 \over |B_n(a)+B^*_n(a')|^2}
	\Bigr]^{1/4}~.
\eqn\local
$$
Its convergence is equivalent to Eq.\orderdiffB~ with $l'=l$, and
$$\eqalign{
	|B_n(a)-B_n(a')|^2&=
	\Bigl( {\mu_n(a) \over g^2(a)}-{\mu_n(a') \over g^2(a')} \Bigr)^2 
\cr    &+
	\Bigl[ {\gamma_n(a)-{a \over N}(\ln a^{1-6\xi})^\cdot \over g^2(a)}
	-{\gamma_n(a')-{a' \over N'}(\ln a'^{1-6\xi})^\cdot \over g^2(a')}
	\Bigr]^2~.
}\eqn\localII
$$
Under Eqs.\behaviormu and \behaviorgamma~, for the decoherence factor, 
Eq.\local~, to be finite, we obtain the constraint 
for rescaling, $g(a)$, and the curvature coupling constant:
$$\eqalign{
	&g(a)=\hbox{constant}~,
\cr
	&\xi={1 \over 6}~.
\cr}\eqn\constraint
$$
%%%%%%%%%%%%%%%%%%%%%%%%%%%%%%%%%%%%%%%%%%%%%%%%%
\chapter{Summary and Discussion}
%%%%%%%%%%%%%%%%%%%%%%%%%%%%%%%%%%%%%%%%%%%%%%%%%
In the previous section we obtained the constraints on 
the rescaling $g(a)$ and the curvature coupling constant 
to make decoherence moderate
{\it irrespective of the evolution of the scale factor}.
Therefore, if the constraints \constraint~ do not hold,
we cannot obtain moderate decoherence, even with 
adiabaticity of the evolution of the scale factor, 
except for in a static region.
Even though we choose the initial condition of 
the state of the scalar field to be a physically reasonable state 
leading to finite particle production,
we have perfect decoherence unless the constraints \constraint~ hold.
So the divergence of the decoherence factor is not attributed to
the lack of adiabaticity of the zeroth order WKB solution,
but to {\it system/environment} coupling and {\it system/environment} 
splitting.
The former is of dynamical origin, but the latter is simply of 
kinematical origin.

Our result, Eq.\constraint~, is understood as follows.
Because we deal with an environment with 
an infinite number of degrees of freedom,
for the decoherence factor not to vanish, 
both $\chi_l(a)$ and $\chi_{l'}(a')$ 
must belong to the same Hilbert space.
In other words, the states of almost all modes in $\chi_l(a)$ should be 
nearly equal to the corresponding modes in $\chi_{l'}(a')$.
Let us consider the case that $|I_{l,l'}(a,a')|=1$, so that
the scalar field essentially has no coupling to the scale factor.
This is the case when we choose $g(a)=1$ 
for the massless conformal coupled scalar.
Unless we choose $g(a)=1$, the effective frequency of each mode is 
$\Omega_{eff}=k/g^4(a)$ and the scale factor strongly couples to modes
irrespective of $k$, $\partial_a \Omega_{eff}/\Omega_{eff}\sim O(k^0)$.
Even if we choose $g(a)=1$, in the case of non-conformal coupling,
we have momentum-momentum coupling, and this gives 
not only $a$-dependence but also direct $l$-dependence 
through $\partial_a S_l$ to the covariance, $B_{nl}(a)$, of 
the wave functions of all modes.
Therefore, the choice of $g(a) \ne 1$ and non-conformal coupling
induce certain differences to all modes, so that the overlap integral
vanishes.
Meanwhile, the mass term is ineffective for large $k$ modes 
and it has no constraint.

We calculate the quantities among a wider class 
in discussing decoherence than in ordinally discussing field theory 
on a curved space.
On the decoherence argument, we calculate the overlap integral 
between the states with the same initial condition 
{\it at different times and in different universes}.
On the other hand, for example, particle creation is calculated by
making a comparison between different states {\it at the same time 
and in the same universe}. 
In usual arguments for field theory on a curved space,
rescaling $g(a)$ and $\xi$ are ineffective.
For example, in discussing that two particle models belong to 
the same Hilbert space, we, in the Schr\"odinger picture, 
calculate the decoherence functional
for $l'=l$(same universe) and $a'=a$(same time), 
but for different particle models($\{u_n\}$ and $\{v_n\}$):
$$
	|I(a,a)|=\Prod n \Bigl[ 1-
	{|B_n(a)-B'_n(a)|^2 \over |B_n(a)+B'^*_n(a)|^2}
	\Bigr]^{1/4}~,
\eqn\specialDF
$$
where $B_n$ and $B'_n$ are written as
$$\eqalignno{
	B_n(a)&=-{i a \over N g^2(a)}
	\Bigl[ \ln u^*_n(a)+\ln a^{1-6\xi} \Bigr]^\cdot~,
&\eqnalign\modeu\cr
	B'_n(a)&=-{i a \over N g^2(a)}
	\Bigl[ \ln v^*_n(a)+\ln a^{1-6\xi} \Bigr]^\cdot~.
&\eqnalign\modev\cr
}$$
Therefore we obtain the well known result
$$\eqalign{
	|I(a,a)|&=\Prod n \Bigl[ 1-
	{|\dot u^* v^*-u^* \dot v^*|^2 \over |\dot u^* v-u^* \dot v|^2}
	\Bigr]^{1/4}
\cr
	&=\Prod n \Bigl[ 1-{|\beta_n|^2 \over |\alpha_n|^2} \Bigr]^{1/4}
	=\Prod n |\alpha_n|^{-1/2}~,
}\eqn\specialDFII
$$
where $v_n=\alpha_n u_n + \beta_n u^*_n$ and the index $l$ is omitted.
Thus by requiring Eq.\specialDFII~ to be convergent, 
we obtain the constraints only on mode functions, 
and these are nothing more than Eqs.\behaviormu~ 
and \behaviorgamma~.

In an anisotropic universe, even if we choose $g(a)=1$, modes tightly
couple to geometry irrespective of $k$, 
$\partial \Omega_n/\Omega_n=[\partial (e^{2\beta})_{ij}k^i k^j+ ...]
/[(e^{2\beta})_{ij}k^i k^j+ ...] \sim O(k^0)$.
We believe that in anisotropic case, the decoherence factor must vanish.
Further, it is known that there is no particle model which leads
to finite particle production.
\refmark\fulling
This may make the WKB ansatz, Eq.\WKB~, invalid.

As described in the Introduction, we have the arbitrariness of defining
the reduced density matrix, $\ie$, one of {\it system/environment}
splitting.
At least in isotropic case, the arbitrariness is important to obtain 
the moderate decoherence that is necessary 
for the classical correlation criterion to hold.
By requiring moderate decoherence, we can obtain a 
{\it system/environment} splitting, $g(a)=\hbox{constant}$, which
is characterized by no coupling between the geometry and scalar field 
in the large $k$ limit.

If we really need moderate decoherence, it may fix a part of 
the arbitrariness of the {\it system/environment} splitting in a sense.
%%%%%%%%%%%%%%%%%%%%%%%%%%%%%%%%%%%%%%%%%%%%%%%%%%%%%%%%%%%%%%%%%%
\ack
%%%%%%%%%%%%%%%%%%%%%%%%%%%%%%%%%%%%%%%%%%%%%%%%%%%%%%%%%%%%%%%%%%
{The author is grateful to Professor A.Hosoya and
Professor H.Ishihara for comments and careful reading of the manuscript.
The author would like to thanks the Japan Society for the Promotion of 
Science for financial support.
This work was supported in part by the Japanese Grant-in-Aid
for Scientific Research from the Ministry of 
Education, Science and Culture.
}

\refout
\end